\documentclass[twocolumn]{openjournal}
\usepackage{epsfig}
\usepackage{amsmath}
\usepackage{hyperref}
\hypersetup{
 colorlinks,
citecolor = [rgb]{0,0.5,0} }
\usepackage{amsmath}
\def\gtwid{\mathrel{\raise.3ex\hbox{$>$\kern-.75em\lower1ex\hbox{$\sim$}}}}
\def\ltwid{\mathrel{\raise.3ex\hbox{$<$\kern-.75em\lower1ex\hbox{$\sim$}}}}
\def\\{\hfil\break}

\def\eg{{\it e.g.}}

\def\sun{\hbox{$\odot$}}
\def\lesssim{\mathrel{\hbox{\rlap{\hbox{\lower2pt\hbox{$\sim$}}}\raise2pt\hbox{$<$}}}}
\def\gtrsim{\mathrel{\hbox{\rlap{\hbox{\lower2pt\hbox{$\sim$}}}\raise2pt\hbox{$>$}}}}

\def\arcdeg{\hbox{$^\circ$}}

\newcommand{\mamo}[1]{\mbox{$#1$}}
\newcommand{\unit}[1]{\ifmmode \:\mbox{\rm #1}\else \mbox{#1}\fi}
\newcommand{\mone}{\mamo{^{-1}}}

\newcommand{\mpc}{\unit{Mpc}}

\newcommand{\hmpc}{\mamo{h\mone}\mpc}

\begin{document}

\title{The Origins of the Bulk flow}

\author{Richard Watkins, Trajan Clark}
\email{rwatkins@willamette.edu}
\email{taclark@willamette.edu}
\affiliation{Department of Physics, Willamette University, Salem, OR 97301, USA}

\author{Hume A. Feldman}
\email{feldman@ku.edu}
\affiliation{Department of Physics \& Astronomy, University of Kansas, Lawrence, KS 66045, USA}

\begin{abstract}
We analyze the origin of the large-scale bulk flow using the CosmicFlows-4 (CF4) peculiar-velocity catalog.  We decompose the observed motions into internal components, generated by mass fluctuations within $200 h^{-1}$Mpc, and external ones arising from structures beyond this volume. A weighted-average technique is developed to test the model’s self-consistency while minimizing the impact of non-Gaussian distance errors. The CF4 velocities show excellent agreement with the predicted internal field, yielding $\beta = 0.31 \pm 0.01$.  We also determine that the value of the Hubble constant that should be used for calculating peculiar velocities from the CF4 to be $H_0 = 75.9 \pm 0.1$ km s$^{-1}$ Mpc$^{-1}$, consistent with CF4 calibrations. Using the minimum-variance formalism, we further separate the bulk flow into its internal and external contributions and find that the observed large-scale bulk flow is dominated by sources beyond $200 h^{-1}$Mpc. The amplitude of this externally driven flow increases monotonically with scale,  consistent with the influence of a distant, massive overdensity. We model this overdensity and determine its most likely mass and distance.  Our results challenge the commonly made assumption that the flow in our local volume due to external mass concentrations can be modeled as being spatially uniform.   
\end{abstract}
\
\section{Introduction}
\label{sec:intro}

The study of large-scale peculiar velocities - departures from the uniform Hubble expansion \citep{Hub1929}  - provides one of the most direct dynamical probes of the large-scale matter distribution in the Universe. On sufficiently large scales, where density perturbations remain in the linear regime, peculiar velocities trace the underlying gravitational potential and thus offer a powerful, independent test of cosmological models \citep{Zwicky1933,RubFor70,ZelSun1980}. Averaged over volumes hundreds of Mpc in radius, these motions combine to form what is known as the bulk flow, a coherent motion of the volume, traced by galaxies within it, relative to the cosmic microwave background (CMB) rest frame \citep[\eg][]{JacBraCiarDav92,Wil94,StrWil95}. Many groups have compiled peculiar velocity catalogs \citep[\eg][]{RubFor70,RubThoForRob76,DreFabBurDav87,LP94,RiePreKir95,FelWat98,ZarBerdaC01,KasAtrKoc08,SprMagCol14,TulCouDol13,CF3} and analyzed them \citep[\eg][]{FelWat94,WatFel95,NusDav95,FelWat98,FelWatMelCha03,HuiGree06,SarFelWat07,WatFel07,FelWat08,Tul08,AbaErd09,WatFelHud09,FelWatHud10,RatKovItz11,DavNusMas11,NusDav11,NusBraDav11,AgaFelWat12,MacFelFerJaf12,TurHudFel12,Nusser14,CarTurLav15,WatFel15a,WatFel15,Nusser16,HelNusFeiBil17,PeeWatFel18,WanRooFelFel2018,BorSupHud2020,SaiColMag2020,StadeJBor2021,MazBanKro23,WanFelWat26} in the past four decades.

In the standard $\Lambda$CDM framework \citep[$\Lambda$CDM][]{PlanckCosPar16}, large-scale motions are directly related to the growth of density fluctuations through gravitational instability.  The assumptions of homogeneity and isotropy imply that coherent motions should diminish with increasing scale;  the standard cosmological model predicts that the root-mean-square amplitude of the bulk flow should be only $\sim$100 km s$^{-1}$ for a volume of $200 h^{-1}$ Mpc; however, several observational analyses over the past two decades have reported bulk flows whose amplitudes appear to exceed these expectations \citep[\eg][]{KasAtrKoc08,WatFelHud09,FelWatHud10,MaGorFel11,MacFelFerJaf12,WatFel15}.  
Such findings, if confirmed, could signal unaccounted-for systematics in peculiar velocity measurements, larger-than-expected density inhomogeneities beyond current survey limits, or even hints of new physics affecting the growth of structure and the homogeneity of the Universe \citep[\eg][]{HasBanKro2020,SecVonRam22}.  

The significance of understanding bulk flows extends beyond local dynamics. Measurements of large-scale velocity coherence test the assumptions of statistical isotropy and homogeneity that underpin $\Lambda$CDM cosmology. They also provide an independent cross-check of parameters such as the growth rate of structure, the bias factor between galaxies and mass, and the normalization of the matter power spectrum, $\sigma_8$. Persistent discrepancies between predicted and observed bulk flows could signal new physics, such as scale-dependent growth, non-trivial dark energy dynamics, or modifications to general relativity on cosmological scales.
For an exhaustive discussion of observational evidence for tensions with $\Lambda$CDM see \citet{CosmoVerse25}.

Recent years have seen significant advances in both data and methodology. The CosmicFlows program \citep[\eg,][]{CF4-full} now provides distance and velocity information for tens of thousands of galaxies, dramatically expanding the reach and statistical power of peculiar-velocity studies. Simultaneously, improved reconstructions of the local density and velocity fields from redshift surveys \citep{CarTurLav15,LilNus21,McaJasAta25}; have enabled more robust comparisons between observed velocities and those predicted by the observed matter distribution. Nonetheless, uncertainties remain in the relative contributions of internal flows, driven by structures within the observed volume, and external flows, arising from mass fluctuations beyond the survey boundary. Disentangling these components is essential for identifying the physical origin of the measured bulk flow and for testing the consistency of the $\Lambda$CDM paradigm on the largest accessible scales. CF4 allows a detailed comparison between observed and predicted velocity fields, enabling the decomposition of the total flow into internal and external contributions.

In this work, we present a new analysis designed to clarify the origin of the bulk flow reported in \cite{WatAllBra23} (hereafter W23) calculated using the most recent CosmicFlows-4 \citep[CF4,][]{CF4-full}  catalog.  They found that the bulk flow in a volume of radius 200$h^{-1}$Mpc has a magnitude of $419\pm 36$ km/s in a direction of Galactic latitude $b = -8^\circ \pm 4^\circ$ and longitude $l = 298^{\circ} \pm 5^\circ$.  They estimated that a bulk flow of this magnitude is in over 4$\sigma$ tension with the standard cosmological model.  

An important concern regarding the finding of a larger than expected bulk flow is that it arises from systematics in the data.  This concern is somewhat assuaged by the fact that W23 found that when they removed either Tully Fisher derived distances or Fundamental Plane distances (the two largest components of the CF4 catalog) the bulk flow remains essentially unchanged.  This makes it unlikely that a systematic in a specific catalog or a distance indicator is the cause of the large bulk flow.   

Also concerning is that  \citet{StiDesLav26,StiDesDev2026} (hereafter S26) claim that large-scale flows are consistent with the standard cosmological model, seemingly in conflict with the findings of W23.  However, these two groups use very different techniques, so it is difficult to make a direct comparison between their results.   First,  unlike S26, W23 use a method that accounts for the fact that peculiar velocity information is concentrated nearby, where uncertainties are smaller and where more objects are measured.  If this concentration of information is not accounted for, results will tend to be dominated by the nearby volume, where peculiar velocities are consistent with the standard cosmological model, as shown by W23.  Second, S26 model the externally sourced flow by a constant dipole or a radially dependent dipole.  As we discuss in Section~\ref{sec:bf}, the flow due to external mass concentrations is not well described by a dipole flow.   Finally, S26 use only the Tully Fisher portion of the CF4 catalog, less than a third of the total group catalog.  This means that the S26 analysis cannot make as stringent test of the standard cosmological model as W23, which uses the entire group catalog \citep[see also][]{Stiskalek2025,stiskalek2026}.  Overall,  more research needs to be done to resolve the differences between S26 and W23.  The large quantity of peculiar velocity data that should be available within the next few years should also help clarify the situation.   

\citet{CarTurLav15} used matter density information from the 2M++  redshift survey \citep{HucMacLuc12} to create a model of the peculiar velocity field originating from mass concentrations within $200h^{-1}$ Mpc.  
\citet{McaJasAta25} have recently improved this model using the same redshift data but including simulations and utilizing Bayesian forward modeling.  For comparison we will employ both of these models separately.  For each model, we use the predicted peculiar velocities to separate the observed peculiar velocities into internal contributions, originating from mass concentrations within $200h^{-1}$ Mpc,  and external contributions, originating from beyond this distance, testing the self-consistency of the model through averaged velocity comparisons. 
This approach mitigates the non-Gaussian effects of distance uncertainties and provides a direct, data-driven way to assess whether the observed flow is compatible with the expected velocity field generated by the local mass distribution.

The remainder of this paper is structured as follows. Section \ref{sec:CF4} describes the CF4 dataset and our processing of the distance and velocity measurements. Section \ref{sec:theory} outlines the theoretical framework used to model internal and external velocity contributions. Section \ref{sec:avg} presents our statistical analysis and tests of model consistency, while Section \ref{sec:bf} discusses the implications of our results for the origin of the observed bulk flow and for potential deviations from $\Lambda$CDM predictions. In Section \ref{sec:model} we model the effect of an external mass concentration on the peculiar velocities in the survey. Section \ref{sec:discussion} summarizes our results and conclusions.

\section{The CosmicFlows4 catalog}
\label{sec:CF4}

The CosmicFlows4 (CF4) catalog \citep{CF4-full} is a compendium of nearly all available cosmological distance measurements.  We use the group version of this catalog.  It gives distance moduli $\mu$ with uncertainties $\sigma_\mu$, redshift $cz$, and location on the sky for over 38,000 galaxies and galaxy groups.  Given that the velocity model only applies out to distances of 200$h^{-1}$Mpc, in this paper we will only use the 23,441 objects with $cz < $ 20,000 km/s from the CF4.  Note that we use redshift rather than distance modulus to locate objects in space as it is more accurate, especially for faraway galaxies.   

In order to calculate peculiar velocities, we first exponentiate distance moduli to get distances.  This process is known to introduce bias into our distances.  However, this bias can be calculated and corrected for \citep[see \eg,][]{WatFel15a}.  Following \citet{WatAllBra23} we calculate corrected distances as
\begin{equation}
d_c = 10^{\frac{\mu}{5} - 5}\ \exp\left(-{(\kappa\sigma_\mu)^2 /2}\right)/(1+z),
\label{eq:dc}
\end{equation}
where the factor of $(1+z)$ converts from luminosity distance to comoving distance, with uncertainties given by
\begin{equation}
\sigma_{c} \approx \kappa\sigma_\mu cz/H_0,
\label{eq:unc}
\end{equation}
where $\kappa=\ln(10)/5$.  

At the distances of objects in the CF4 catalog, the difference between luminosity distances and comoving distances, as well as deviations from the linear Hubble's law, have a significant effect on calculations of peculiar velocities.  A common approach is to account for both of these effects is by using the modified redshift $z_{mod}$, calculated as a Taylor series in $z$, \citep[see \eg,][]{Visser04,RieMacCas09,DavScr14,WatFel15a}.
\begin{align}
z_{mod} = z & \left( 1+ \frac{1}{2}(1-q_o) z \right. \nonumber \\
	& \left. - \frac{1}{6}( 1- q_o - 3q_o^2 + j_o )z^2  + \mathcal{O}(z^3)\right),
\end{align}
where $q_o$ is the deceleration parameter, $j_o$ is the jerk, and we have assumed a spatially flat Universe.   The parameters $q_o$ and $j_o$ can be determined empirically \citep[see \eg,][]{RieStrCas07,FreRidKra17}. 

In terms of the modified redshift, the unbiased luminosity distances can be used to calculate unbiased peculiar velocities through the formula
\begin{equation}
v_{c} = (cz_{mod} - H_0 d_c)/(1+z).
\label{eq:vc}
\end{equation}
We note that while the peculiar velocity estimates given here are unbiased, in that averages over velocities will give the correct result,  their probability distributions are skewed and thus nonGaussian.  This point warrants further discussion.  Velocity estimates are unbiased which tells us that the errors in the velocities will average to zero, even if the distributions of errors is nonGaussian.   However, likelihood analyses typically involve the squares of velocity errors.  Squaring values that are drawn from a skewed distribution makes their distribution even more skewed and can bias results.  Velocities with skewed error distributions can give much different values of $(v-v_{model})^2$ than velocities with Gaussian errors, even for the same level of uncertainties.   While there are velocity estimators that result in Gaussian errors \citep[see][]{WatFel15a}, these estimators are not accurate for all distances.    Many researchers choose to work with distance moduli instead of distances or velocities, since distance moduli have Gaussian errors \citep[see \eg,][]{SprHonSta16}.  However, this method involves predicting distance moduli from redshifts and these distance moduli will have nonGaussian errors due to peculiar velocities.   



\section{Theory}
\label{sec:theory}

We begin by assuming that the CF4 peculiar velocities $v_i$ are the sum of a contribution from the mass distribution inside the region of radius $200h^{-1}$Mpc, which we will call $v_{int,i}$, a contribution from the mass distribution outside of this region, which we will call $v_{ext,i}$, measurement noise $\delta_i$, and an additional contribution $\delta_{*,i}$ due to galaxies  not being perfect tracers of the velocity field.  Thus we have
\begin{equation}
v_i = v_{int,i} + v_{ext,i}+\delta_i + \delta_{*,i}.
\label{eq:vi}
\end{equation}

We use the velocity field models of \citet{CarTurLav15} and \citet{McaJasAta25} separately to calculate $v_{int,i}$ for the objects in the CF4 that are within 200$h^{-1}$Mpc.  We will assume that $\delta_{*,i}$ are drawn from a Gaussian distribution of width $\sigma_*$, which we will take to be $150$km/s \citep[see \eg,][]{TurHudFel12}.  Gaussian measurement errors in the distance modulus $\mu$ are translated into non-Gaussian distributed errors when moduli are exponentiated to obtain distance and subsequently velocity errors $\delta_i$.    However, we can eliminate bias from estimates of $v_i$, which ensures that the $\delta_i$ average to zero (see Sec.~\ref{sec:CF4} above).  
Note that while it is possible to use a velocity estimator that has Gaussian distributed errors \citep[see, \eg,][]{WatFel15a}, these estimators are not accurate at small redshift $cz$, which is where much of the information in our data lies.



\section{Testing theory with averages}
\label{sec:avg}

As we have discussed above, it is possible to eliminate biases in \textit{averages} even when error distributions are non-Gaussian.  Here we will focus on using averages of radial velocities as a tool to study Eq.~\ref{eq:vi}.  
%

As a test of our hypothesis, we average the CF4 velocities in bins based on their values of $v_{int}$.  Since the other terms in the expression given in Eq.~\ref{eq:vi} should  be uncorrelated with $v_{int}$, we should ideally have $\langle v\rangle = v_{int}$, where the angle brackets signify an average over the peculiar velocities in an individual $v_{int}$ bin.  Since objects in a given bin have a large range of velocity uncertainties, we do a weighted average over the objects in each bin, using weights $w_n= (\sigma_n^2 + \sigma_*^2)^{-1}$, where the index $n$ references the objects in a particular bin.  In order to have similar uncertainties for the averages in each bin, we choose bins to have equal numbers of $N$ objects.  In this paper we used $N=$3,000 (the results are fairly independent of this choice).  We calculate the $v_{int}$ value to assign to each bin using a weighted average over the $v_{int,n}$ values of the objects in the bin using the same weights as for $\langle v\rangle$.  

Ideally, our averages would fall on a line of slope unity and a $y$ intercept of zero, indicating the validity of the velocity model given in Eq.~\ref{eq:vi}.  There are two parameters that we can adjust in our model to achieve this result.  
\begin{figure}
\centering
\includegraphics[scale=0.5]{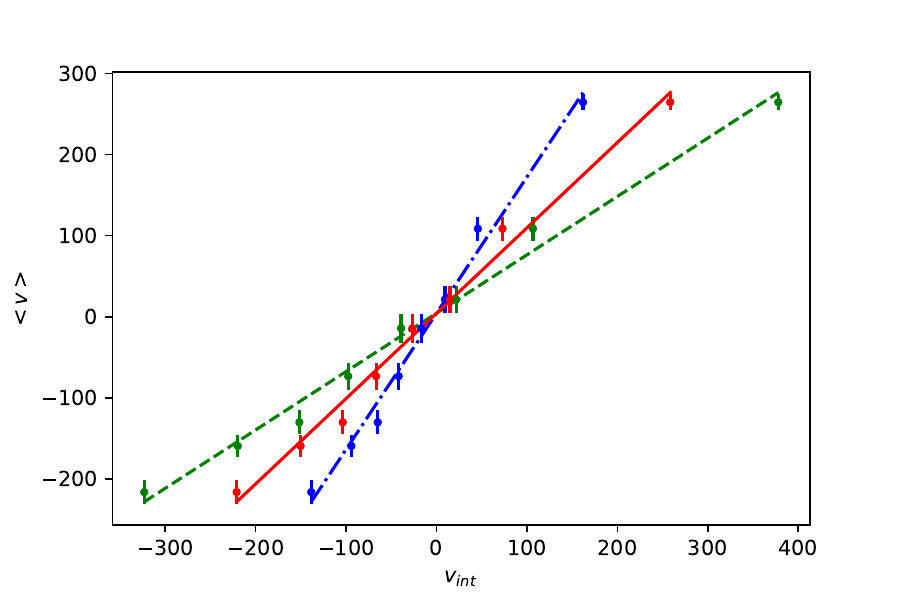}
\caption{Weighted averages of the measured peculiar velocities in bins of $v_{int}$ from \citet{CarTurLav15} for the best fit value of $H_0$ but varying values of $\beta$.  Each bin contains 3,000 objects. The blue points and dash-dot slope correspond to $\beta= 0.43$, the red points and solid slope correspond to $\beta=0.31$ (the best fit value), and the green points and dashed slope correspond to $\beta=0.19$   Velocity of the bin is determined by the weighted average of $v_{int}$ values of the objects in the bin.  The red solid line in the figure has a slope of one. } 
\label{fig:betavar}
\end{figure}
\begin{figure}
\centering
\includegraphics[scale=0.5]{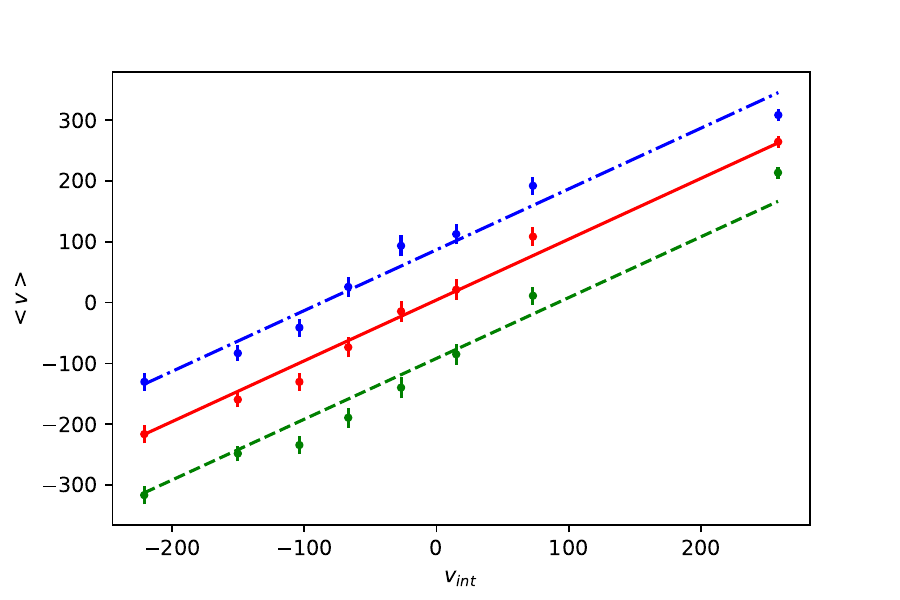}
\caption{Weighted averages of the measured peculiar velocities in bins of $v_{int}$ from \citet{CarTurLav15} for the best fit value of $\beta$ but varying values of $H_0$.  Each bin contains 3,000 objects.  The green points and dashed slope correspond to $H_0=  74$km/s/Mpc, the red points and solid line correspond to $H_0=75.8$km/s/Mpc (the best fit value), and the blue points and dash-dot slope correspond to $H_0=78$km/s/Mpc   Velocity of the bin is determined by the weighted average of $v_{int}$ values of the objects in the bin.  All lines in the figure have a slope of one. }  
\label{fig:hvar}
\end{figure}

First, the velocity scale $\beta= \Omega_m^{0.55}/b$, where $\Omega_m$ is the mass density parameter and $b$ is the bias parameter,  determines the growth rate of structure and hence the amplitude of the internal velocities, \citep[see \eg,][]{Linder05,SelMakMan05,DavNusMas11,SprHonSta16}.  Changing $\beta$ multiplies the internal velocities by a factor, thus changing the slope of the relationship with the bin averages.  This is shown in Fig.~\ref{fig:betavar}, where we plot the averages for three different values of $\beta$.  

Second, we can take the value of the Hubble constant $H_0$ used to calculate the CF4 velocities to be a free parameter in our analysis.  Increasing $H_0$ adds an amount proportional to distance to all radial velocities, and thus shifts the line of averages upward.  Similarly, a decrease in $H_0$ will shift the line of averages downward.  This is shown in Fig.~\ref{fig:hvar}, where we plot the averages for three different values of $H_0$. 
Since these two parameters change the line of averages in very different ways 
they are essentially uncorrelated, and we can determine the best fit values of $\beta$ and $H_0$ independently by minimizing the difference between our averages and a line of slope unity and vanishing y-intercept.    

\begin{figure}
\centering
\includegraphics[scale=0.5]{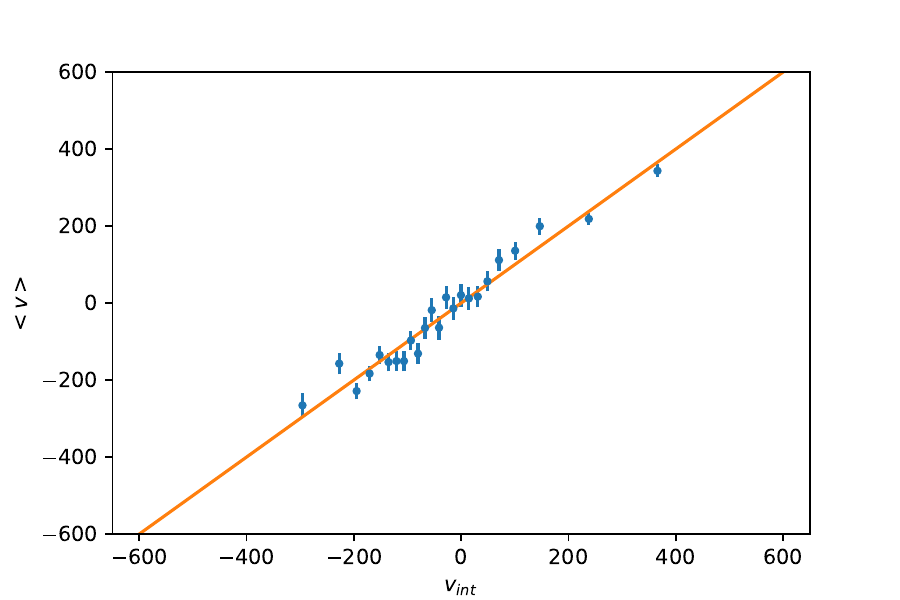}
\caption{Weighted averages of the measured peculiar velocities in bins of $v_{int}$ from \citet{CarTurLav15} with their uncertainties, for the best fit values of $\beta$ and $H_0$ and with $N=$1,000 objects per bin.   The velocity of the bin is determined by the weighted average of $v_{int}$ values of the objects in the bin.  The line in the figure has a slope of one (the same as the red solid line in Fig.~\ref{fig:hvar}). }
\label{fig:v_int_avg}
\end{figure}
\begin{figure}
\centering
\includegraphics[scale=0.5]{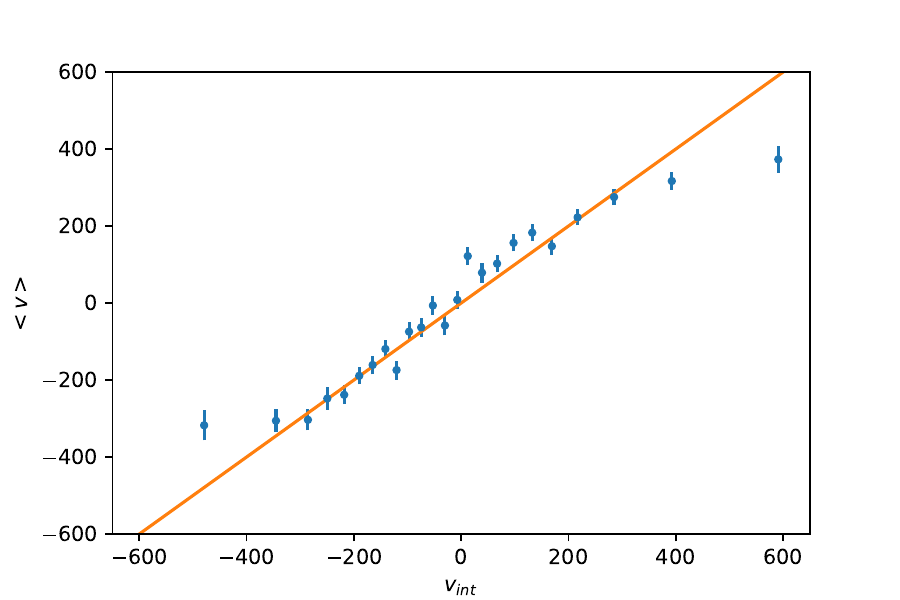}
\caption{Same as Fig.~\ref{fig:v_int_avg} but with data from \citet{McaJasAta25} with best value of $H_0$ (as mentioned in the text, $\beta$ is not a parameter that can be adjusted). The line in the figure has a slope of one (the same as the red solid line in Fig.~\ref{fig:hvar}). }
\label{fig:v_int_avg-mant}
\end{figure}

Fig.~\ref{fig:v_int_avg} shows the bin averages for the best fit values of $\beta$ and $H_0$ with their uncertainties for the internal velocities given by \citet{CarTurLav15}.  The $v_{int}$ bins contain an equal number of velocities, $N=$1,000.  Clearly this matches a line of slope unity well.  Our analysis gives $\beta = 0.31\pm 0.01$ and $H_0 = 75.9\pm 0.1$km/s/Mpc.  Our value of $\beta$ is quite consistent with that obtained through a comparison of the 6dF and SDSS surveys  to the \citet{CarTurLav15} velocities \citep{SaiColMag}.  Interestingly,  
\citet{Stiskalek2025} determine a somewhat larger value for $\beta$ in conflict with \citet{SaiColMag} for reasons that remain obscure.  We note that our purpose here is not to make a measurement of $\beta$, but rather to determine the best value of $\beta$ to use when using the \citet{CarTurLav15} velocities together with the CF4.  

Our $H_0$ value is consistent with that determined by the calibration of the CF4 \citep{CF4-full},  with somewhat smaller uncertainties.  Note that, as with $\beta$, our purpose here is not to make a measurement of $H_0$.   This estimate of $H_0$ is specifically the value that should be used with the CF4 to calculate peculiar velocities; our analysis does not include an examination of various systematics that would be necessary in a more general estimation of $H_0$.  

Fig.~\ref{fig:v_int_avg-mant} shows the bin averages for the best fit value of $H_0$ with their uncertainties for the internal velocities given by \citet{McaJasAta25}.  Due to their use of simulations, the bias is taken into account implicitly in their calculations and $\beta$ is not a parameter that can be adjusted.  We obtain a best fit value of $H_0= 75.84\pm 0.08$km/s, which is in fair agreement with the value obtained using \citet{CarTurLav15} velocities.  Comparing  Fig.~\ref{fig:v_int_avg} and Fig.~\ref{fig:v_int_avg-mant} we note that the \citet{McaJasAta25} model produces a wider range of velocities.  While the averages are in good agreement for the smaller velocity bins, we see that the averages underestimate the velocity for the largest velocity bins.  \citet{CarTurLav15} doesn't give these large velocities and we don't see this effect in their work.  We hypothesize that while \citet{CarTurLav15} relies on linear theory, the model of \citet{McaJasAta25} incorporates nonlinear effects; in particular, the large velocities associated with infall of galaxies into clusters.  
We are using the groups version of the CF4 catalog, which combines the peculiar velocities of galaxies in a group or cluster to give a single peculiar velocity for that object.  Thus we would not expect to see these large galaxy infall velocities in the CF4 data.  This explains why averages in these large velocity bins are underestimated.  


The  velocities from \citet{McaJasAta25} already assume a fixed $\beta$ and provide a good match with the CF4 bin averages without the need for any additional scaling.  The $H_0$ value obtained using both models is consistent with that determined by the calibration of the CF4 \citep{CF4-full},  with somewhat smaller uncertainties, and consistent with each other to a high precision.  Note that this estimate of $H_0$ is specifically the value that should be used with the CF4 to calculate peculiar velocities; our analysis does not include an examination of various systematics that would be necessary in a general estimation of $H_0$.  

One might be concerned that this analysis ignores radial inflows or outflows, which can mimic variations in the Hubble constant.  However, since we weigh by uncertainty, which generally grows with distance, our weighted averages are dominated by nearby objects.  Since any radial flow must vanish at the origin, it is unlikely that radial flows would have a significant effect on our analysis.  A radial flow near our position would also imply that we are at the center of either a void or an overdensity, which is both unlikely and not supported by any observational evidence we are aware of.   
\begin{figure}
\centering
\includegraphics[scale=0.5]{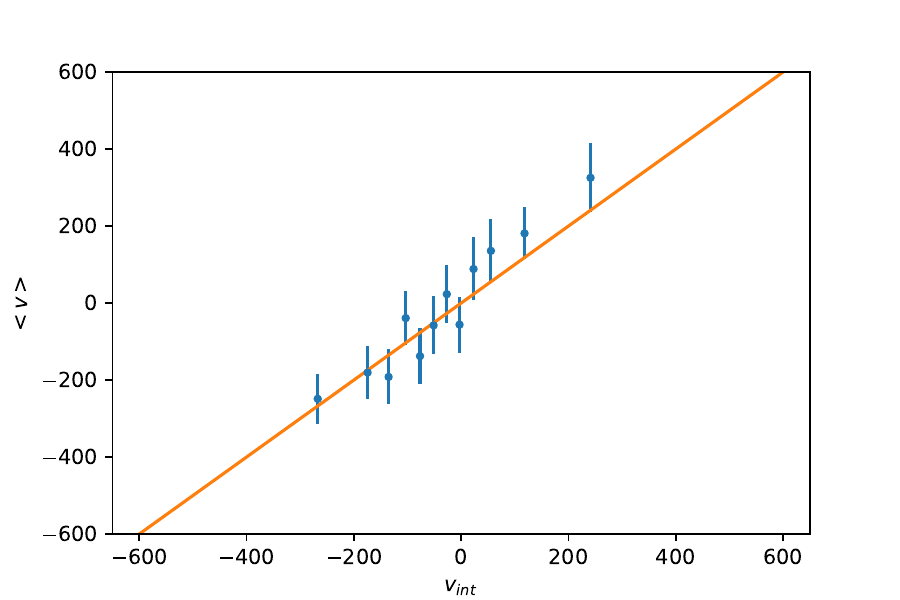}
\caption{Weighted averages of the measured peculiar velocities in bins of $v_{int}$ from \citet{CarTurLav15} and their uncertainties, for objects with $r> 100h^{-1}$Mpc and using the same best fit values of $\beta$ and $H_0$ used in Fig.~\ref{fig:v_int_avg} and the same $N=$1,000 per bin.   Velocity of the bin is determined by the weighted average of $v_{int}$ values of the objects in the bin.  The line in the figure has a slope of one. } 
\label{fig:v_int_avg-g120}
\end{figure}
\begin{figure}
\centering
\includegraphics[scale=0.5]{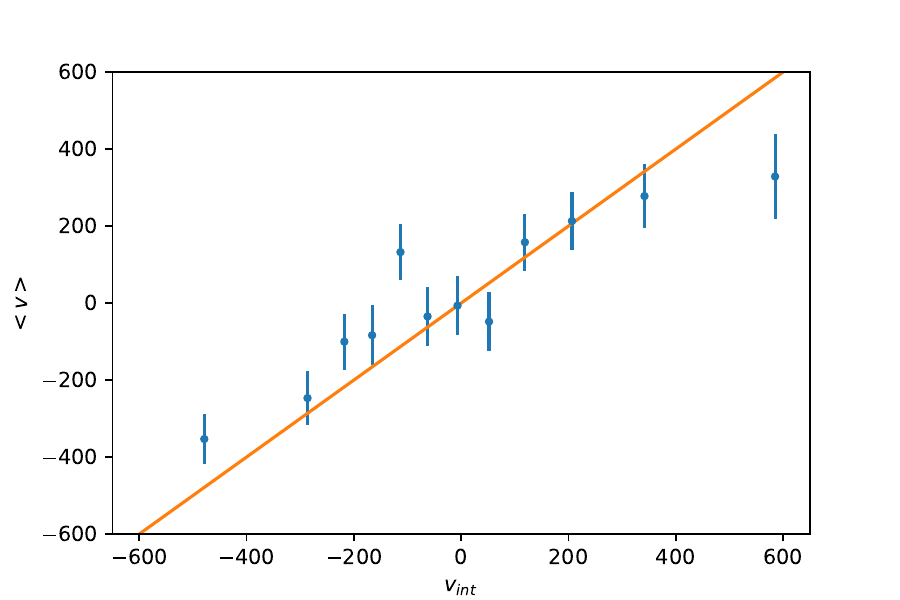}
\caption{Same as Fig.~\ref{fig:v_int_avg-g120} but with data from \citet{McaJasAta25}. The line in the figure has a slope of one.}
\label{fig:v_int_avg-g120-mant}
\end{figure}

In order to test that Eq.~\ref{eq:vi} holds equally well for deeper objects, we can institute a minimum distance cutoff $R_{min}$ and only average only objects beyond the cutoff. In Fig.~\ref{fig:v_int_avg-g120}  (Fig.~\ref{fig:v_int_avg-g120-mant}) we show the same weighted averages but this time only for objects at distances $r> 120h^{-1}$Mpc.  Here we use the same best fit values of $\beta$ and $H_0$ as in Fig.~\ref{fig:v_int_avg} (Fig.~\ref{fig:v_int_avg-mant} with only best fit $H_0$).   Farther away objects apparently average to the same line as nearby objects, suggesting that radial flows are not having a significant effect on our averages.

Figs.~\ref{fig:v_int_avg} and ~\ref{fig:v_int_avg-g120}  show a good match between the averages and the $v_{int}$ bin values for our best fit $\beta$ and $H_0$, thus supporting our velocity model (the same holds for Figs.~\ref{fig:v_int_avg-mant} and ~\ref{fig:v_int_avg-g120-mant} with only best fit $H_0$).  The level of agreement is particularly striking given that the internal velocities for both \citet{CarTurLav15} and \citet{McaJasAta25} were calculated from mass distributions inferred from redshift surveys and the CF4 peculiar velocities were calculated using a combination of redshift and distance estimates.  The agreement between them gives us confidence that both methods are accurately estimating velocities, and confirms other studies that make these comparisons using different methods (see \eg, \cite{SaiColMag,StiDesDev25}).

Our method provides a novel way of measuring $\beta$ for the CF4 galaxies and the $H_0$ value that should be used to calculate peculiar velocities for the CF4 very accurately.   It is worth noting that the bulk flow calculated in \citet{WatAllBra23} does not depend on the value of the $H_0$ used.  


\begin{figure}
\centering
\includegraphics[scale=0.5]{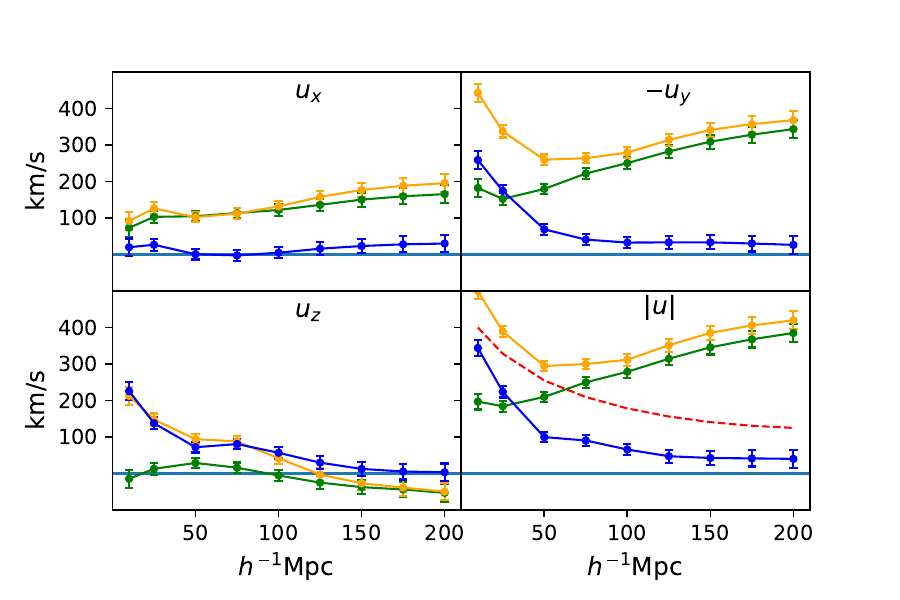}
\caption{The components ($U_x,-U_y,U_z$) of the bulk flow and their magnitude ($|U|$) calculated using the MV method for different radius volumes. The orange line is the total bulk flow.  The blue line is contribution to the bulk flow from velocities generated by mass distributions within 200$h^{-1}$Mpc from the \citet{CarTurLav15} model.  The green line is the bulk flow from velocities generated by mass distributions outside of 200$h^{-1}$Mpc.  }  
\label{fig:bf}
\end{figure}

\section{The Origin of the Bulk Flow}
\label{sec:bf}

Now that we have validated our model for the CF4 peculiar velocities, we can obtain an estimate for the contribution to peculiar velocities from external sources $v_{ext,i}$ by subtracting $v_{int,i}$ from the peculiar velocities $v_i$,
\begin{equation}
{v_{ext,i}} = v_i-  v_{int,i}.
\label{eq:vext}
\end{equation}

\begin{figure}
\centering
\includegraphics[scale=0.5]{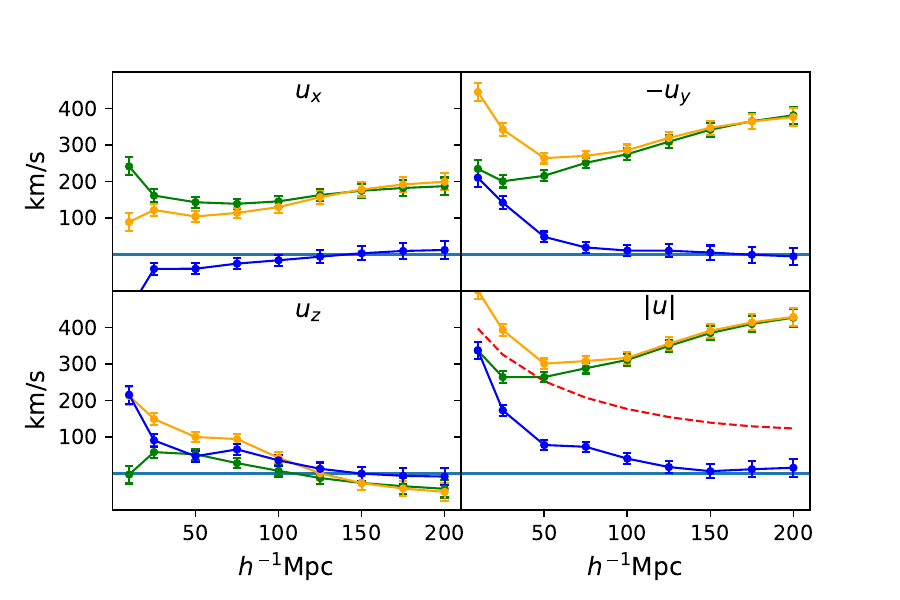}
\caption{Same as Fig.~\ref{fig:bf} but with data from \citet{McaJasAta25}.}
\label{fig:bf-mant}
\end{figure}

Here we consider the bulk flow calculated using the minimum variance (MV) method \citep{WatFelHud09,FelWatHud10,WatFel15,PeeWatFel18,WatAllBra23}.  
In this formalism the bulk flow is calculated through a weighted average of the peculiar velocities, 
$U_j = \sum w_{jk}v_k,$ where $w_{jk}$ determines the contribution of the velocity of the $k^{th}$ galaxy to the $j^{th}$ component of the bulk flow.  We can use Eq.~\ref{eq:vext} to separate out the bulk flow arising from external mass distributions from that due to internal mass distributions; 
\begin{equation}
\begin{split}
U_{ext,j} &= \sum w_{jk}v_{ext,k}\\
U_{int,j} &= \sum w_{jk}v_{int,k}\\
\Rightarrow U_j &= U_{ext,j}+U_{int,j}
\label{eq:U}
\end{split}
\end{equation}

In Fig.~\ref{fig:bf} we show the bulk flow components and magnitudes for the total peculiar velocity as well as for both the contributions from internal (calculated using the \citet{CarTurLav15} model) and external sources.  Fig.~\ref{fig:bf-mant} shows the same thing except using the \citet{McaJasAta25} model.   What is noteworthy in these figures is that even within a radius as small as $50 h^{-1}$Mpc, the bulk flow is dominated by mass distributions located beyond 200$h^{-1}$Mpc.  This is especially striking given that the influence of the Shapley Concentration \citep{Shapley30,BolHel08} is included in $U_{int,j}$.    The two plots tell essentially the same story.

\begin{figure}
\centering
\includegraphics[scale=0.5]{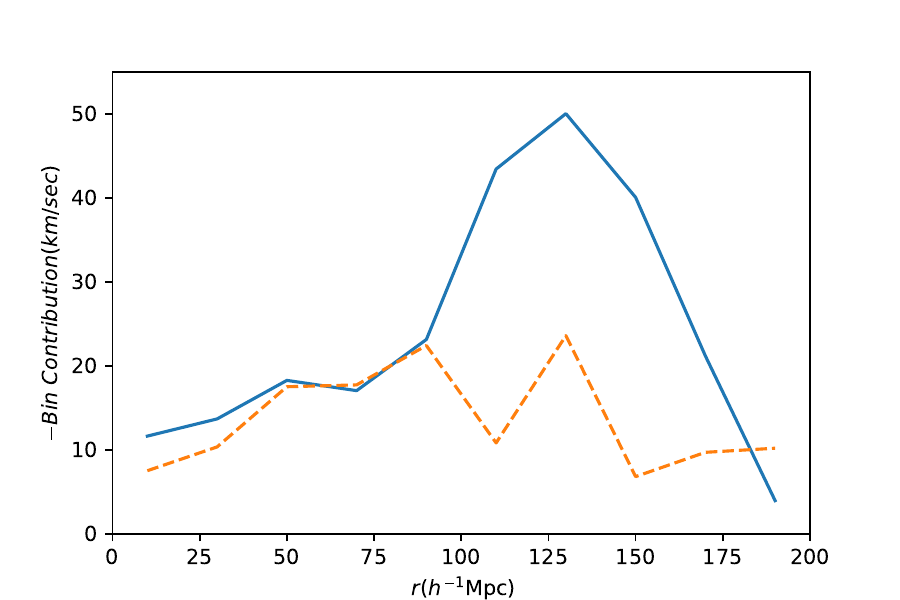}
\caption{ The summed contribution to the bulk flow due to external sources (with the \citet{CarTurLav15} model velocities subtracted out) for a sphere of radius $200 h^{-1}$Mpc in radial  bins of width $20$\hmpc\ and separated into contributions from the hemisphere in the direction of the bulk flow (the blue solid line) and the hemisphere opposite the bulk flow direction (the orange dashed line).     We see that the contributions to the bulk flow are largest on the edge of the volume closest to the bulk flow direction, becoming smaller as we move away from the source of the bulk flow.  }  
\label{fig:contbf}
\end{figure}
\begin{figure}
\centering
\includegraphics[scale=0.5]{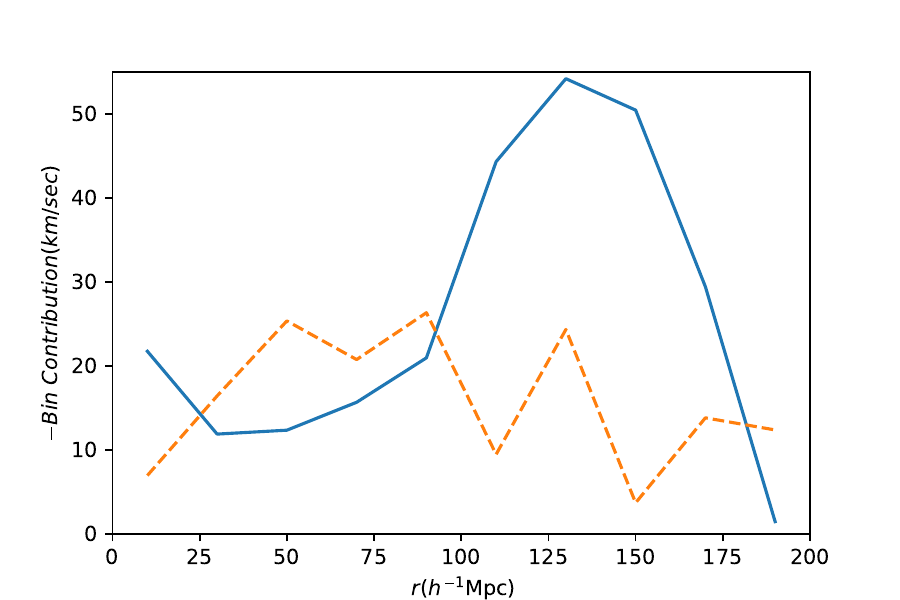}
\caption{Same as Fig.~\ref{fig:contbf} but with data from \citet{McaJasAta25}.}
\label{fig:contbf-mant}
\end{figure}

Furthermore, Figs.~\ref{fig:bf} and \ref{fig:bf-mant} show that the magnitude of the bulk flow from external mass distributions rises monotonically with radius.  The most natural explanation for this is the presence of a substantial mass concentration located in the direction of the bulk flow but lying outside the survey volume. Such a structure would exert its strongest influence near the survey boundary, with the induced flow diminishing toward the Milky Way and beyond. We test this hypothesis by examining the contribution of each galaxy to the externally sourced bulk flow, enabling us to trace the flow pattern throughout the volume.


First, we calculate the contributions to the bulk flow from each object in a sphere of radius $200 h^{-1}$Mpc.  The contribution that the $k^{th}$ galaxy's velocity makes to the  bulk flow is given by $\sum_j \hat U_j w_{jk} v_{ext,k}$, where the index $k$ is not being summed over and $\hat U$ is a unit vector in the direction of the bulk flow.  By summing these contributions in bins we can determine how much each region of the volume is contributing to the bulk flow.    We sum the contributions in radial bins of width $20$\hmpc,  separating contributions into those coming from the hemisphere in the direction of the bulk flow and those coming from the hemisphere opposite the bulk flow direction.   Note that since we only measure the radial component of a galaxy's velocity, objects with position vectors orthogonal to the bulk flow direction cannot make a contribution to the bulk flow.   The result of this calculation for both of our velocity models is shown in Figs.~\ref{fig:contbf} and \ref{fig:contbf-mant}.    Consistent with our earlier hypothesis, the contributions to the bulk flow are largest near the edge of the volume in the bulk-flow direction, falling off with increasing distance from that edge. Note that there are no objects beyond $155h^{-1}$Mpc in the direction of the bulk flow, leading to the contribution falling off at the very edge of the volume.


Upon reflection, it seems reasonable that the influence of mass distributions outside  the survey volume will be largest at the edges of the survey volume, where they are closest to these mass distributions.   However, many papers that attempt to model the velocity field due to external mass distributions  model it to be a constant flow over the volume  
(both \citet{CarTurLav15} and \citet{McaJasAta25} do this).  This assumption seems to be in conflict with the results presented in this paper.  Perhaps even more concerning, this model of a constant flow due to external sources is used in a forward-modeling procedure for calculating Tully-Fisher velocities \citep{BouColSai24a,BouColSai24b,BouColSai25}.  It is important to investigate the impact that the constant external flow model has on the results of this kind of  analysis, and to develop more accurate models for the velocity field in our volume due to external sources.

\section{Modeling an External Mass Concentration}
\label{sec:model}

In this section, we model the effect of an external mass concentration on the peculiar velocities in our survey.  We will assume a mass concentration centered at a distance $R_m$ and galactic longitude and lattitude $l_m$ and $b_m$ respectively.    Following \citet{LynFabBur88}, we take the induced peculiar velocity from the mass concentration to fall off as a power law $1/r^\alpha$ and parametrize the strength of the mass concentration by the peculiar velocity $V_m$ it induces at the location of the Local Group.  Like \citet{LynFabBur88}, we will consider $\alpha=1$, which corresponds to a $1/r^2$ decrease in excess density, but we will also consider $\alpha=2$, which corresponds to a more point-like mass concentration.  

In order to find the best fit parameters $R_m$, $l_m$, $b_m$, and $V_m$, we employ Markov Chain Monte Carlo (MCMC), minimizing the quantity
\begin{equation}
\sum_i \frac{(v_i-v_{im})^2}{\sigma_i^2 + \sigma_*^2},
\end{equation}
where $v_i$ is the measured peculiar velocity with uncertainty $\sigma_i$ of the $i$th object, $v_{im}$ is the model peculiar velocity for that galaxy, and $\sigma_*$ accounts for the fact that galaxies are not perfect tracers of the underlying velocity field.   Following previous research we take $\sigma_*=150$km/sec, although our results do not depend strongly on its precise value.  

Both values of $\alpha$ give almost identical best fit directions for the mass concentration of $l_m= 295^\circ\pm 3^\circ$ and $b_m= -3^\circ\pm 2^\circ$.  This direction is quite consistent with the direction of the bulk flow measured in \cite{WatAllBra23}.  

\begin{figure}
\centering
\includegraphics[scale=0.5]{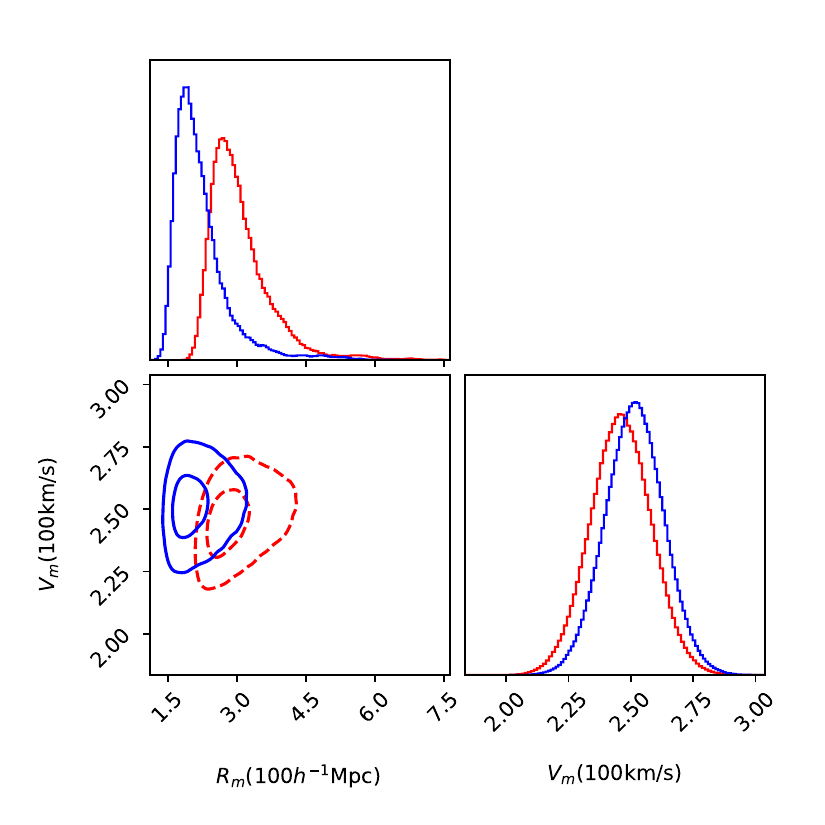}
\caption{ A corner plot showing the likelihood for $R_m$ and $V_m$ for our model.  The solid (blue) contour is for $\alpha=1$ and the dashed (red) contour is for $\alpha=2$.    }  
\label{fig:corner}
\end{figure}

In Fig.~\ref{fig:corner} we show the results of the MCMC for the parameters $R_m$ and $V_m$.  We see that for both values of $\alpha$, the most likely distance for the mass concentration is just outside the 200 \hmpc\ radius volume.  Indeed, for $\alpha=1$ there is a significant likelihood for the mass concentration to be inside the volume.    However, this is not in conflict with our sense of velocities being induced from internal and external sources since the region in question is behind the galactic plane, is very poorly sampled in the 2M++ redshift survey, and so is not accounted for in the internal velocities.  

The mass concentration model does a good job of explaining the larger than expected bulk flow reported in \citep{WatAllBra23}.  Subtracting the velocities mass concentration model from the observed peculiar velocities results in a bulk flow in a region of radius of $200$\hmpc\ of $\vec U= (66  , -153,   -3)$ km/sec for $\alpha=1$ or $(56, -130,    3)$ km/sec for $\alpha=2$ (both in galactic coordinates).  Thus accounting for the mass concentration reduces the bulk flow to a level that is consistent with the cosmological standard model.  

We can estimate the excess mass $\delta M$ in the mass concentration using linear theory.  The continuity equation states that
\begin{equation}
{\bf\nabla}\cdot {\bf v} = -aHf(\Omega_m) \delta,
\end{equation}
where ${\bf v}$ is the velocity field, $a$ is the scale factor, $H$ is the Hubble parameter, $f(\Omega_m)$ is the linear growth rate, and $\delta$ is the density contrast.  Integrating both sides over a sphere of radius $R_m$ centered on the mass concentration and using the divergence theorem on the left hand side gives
\begin{equation}
\oint_{S} \mathbf{v} \cdot d\mathbf{\sigma} = -a H f(\Omega_m) \left(\frac{4\pi R_m^3}{3} \bar\delta_M\right),
\end{equation}
where $S$ is the surface of the sphere and $\bar\delta_M$ is the average density in the mass concentration.  Assuming spherical symmetry allows us to write the left hand side as $-V_m\left (4\pi R_m^2\right)$, where $V_m$ is the magnitude of the infall velocity at the surface of the sphere.  Dividing through by the area and using $a=1$ and $H=H_0$ today gives
\begin{equation}
V_m = H_0 f(\Omega_m) \left(\frac{R_m}{3}{ \bar{\delta}_M}\right).
\end{equation}
Finally, we use the fact that $ \bar{\delta}_M = \frac{\delta M}{\overline M}$, where ${\overline M}$ is the average mass in a sphere of radius R, which can be written in terms of the critical density $\rho_c = \frac{3H_0^2}{8\pi G}$,
\begin{equation}
{\overline M}= \frac{4\pi}{3}R_m^3 \Omega_m \rho_c =  \frac{H_0^2 R_m^3 \Omega_m}{2 G}.
\end{equation}
Putting everything together and rearranging gives
\begin{equation}
\delta M = \frac{3\Omega_m H_0 R_m^2 V_m}{2 f(\Omega_m)G}.
\end{equation}
Taking $\Omega_m=0.3$, $H_0= 100\, h$km/sec/Mpc and noting the $R_m$ is measured in \hmpc, as a rough estimate we get  $\delta M \approx 2\cdot 10^{17}M_\odot h^{-1}$ for $\alpha=1$, and $\delta M \approx 4\cdot 10^{17}M_\odot h^{-1}$ for $\alpha=2$.  

Given that the region where this mass concentration should be is in the zone of avoidance, it is plausible that it has not been seen in large-scale surveys.  In fact, \citet{HollCouKra26} have recently used peculiar velocity and redshift survey information together to estimate the mass of the Vela supercluster, which is also in the zone of avoidance.  Their mass estimate is comparable to that we find to explain the large bulk flow.  The direction and distance to Vela are also roughly consistent with our results.  Clearly this region is deserving of further study.

\section{Discussion}
\label{sec:discussion}

Previous work \citep{WatAllBra23} used the CF4 catalog to estimate the bulk flow for a volume of radius $200 h^{-1}$Mpc and found it to be in significant tension with the cosmological standard model.  In this context, then, it is important to understand the origin of the bulk flow; how much of the bulk flow is from local mass concentrations and how much is caused by sources beyond the survey volume.  In this paper, we have compared the CF4 peculiar velocities to the expected velocities $v_{int}$ given by a model that uses redshift survey information to infer mass concentrations within $200 h^{-1}$Mpc volume.  By averaging velocities in bins of $v_{int}$, we show excellent agreement between the CF4 and both \citet{CarTurLav15} and \citet{McaJasAta25} velocities, suggesting that we can accurately separate the contributions to peculiar velocities from local mass distributions from those due to farther away mass concentrations.  Our averaging also provides a very precise estimate of the growth factor $\beta$ and the value of the Hubble constant $H_0$ that should be used to calculate peculiar velocities from the CF4 distances and redshifts.   

Separating the total bulk flow into the sum of a locally sourced bulk flow and a bulk flow from external sources, we find that the observed bulk flow is primarily due to external mass concentrations.  We observe the external bulk flow to \textit{increase} with increasing radius out to the limit of the current data.  Moreover, by analyzing the contributions to the bulk flow from different regions of the volume, we find that the externally sourced component is strongly non-uniform. The flow is largest in the region closest to the bulk-flow direction and declines steadily with increasing 
distance from this area.  

This is suggestive of a large mass concentration outside the volume in the direction of the bulk flow, and seemingly conflicts with velocities having non-gravitational origins, such as violations of the cosmological principle (see \eg, \cite{BohLukSch2025,  SecVonRam22, SecVonRam21}) 
or tilted universe models \citep{Tur91,Tur92,KasAtrKoc08,MaGorFel11}, where we would expect contributions to the bulk flow to be roughly uniform across the volume.  
There are assertions that suggest that splitting the CMB into half sky maps show variations in $H_0$ of order of $10\%$, which may indicate anisotropy in the Universe \citep[see, \eg,][]{AluCeaChi2022, FerRamFer25}.  \cite{StiDesLav26} examine peculiar velocities and do not see a significant anisotropy in $H_0$; however, they use only a fraction of the full CF4 catalog in their analysis.  

Also relevant are recent studies that argue that in standard general relativity there is faster growth of peculiar velocities than in the quasi-Newtonian approximation commonly employed \citep[for a recent summary see][]{tsagas26}.  Not only could this faster growth help explain the larger than expected bulk flow, the subsequent decay of peculiar velocities during the epoch of accelerated expansion could provide a theoretical explanation for the growth of the bulk flow with radius seen in Figs.~\ref{fig:bf} and \ref{fig:bf-mant}.

It is not surprising that external mass concentrations will induce non-uniform flows in the local volume.  However, models of the velocity field typically assume that external mass distributions induce a uniform flow in the local volume.  The results that we have found here cast doubt on this hypothesis.  More work is required to determine the impact of this assumption and to develop more accurate models of the velocity field.  

Many questions remain regarding how to reconcile the observed bulk flow with the expectations of the standard cosmological model.  Our work showing that the mass concentration responsible for the bulk flow must be further away than $200 h^{-1}$Mpc exacerbates the problem, since the further away the mass concentration is, the larger it must be.   While the mass concentration required to explain the observed bulk flow is unlikely in the standard model, recent observations by the James Webb telescope \citep[see \eg,][]{Gup23} and  studies that suggest the existence of a local (KBC) supervoid \citep[see \eg,][]{HasBanKro20,MazBanKro23,StiDesBan25,BanKal25} may violate the cosmological principle and challenge $\Lambda$CDM and thus affect bulk flow measurements 
\citep[note that][argues against a supervoid]{StiDesBan25}.  

More data in the form of deeper redshift surveys and peculiar velocity studies would help clarify the situation.  In particular, more data in the direction of the bulk flow is crucial.  This region of the sky is close to the galactic plane and relatively understudied.  

Ultimately, whether the observed flow represents an extreme, but rare, fluctuation within $\Lambda$CDM, an incomplete accounting of the large-scale mass distribution, or a sign of physics beyond the cosmological standard model, remain an open question. Upcoming wide-field surveys and precise distance measurements, particularly in regions currently obscured by the Galactic plane, will be decisive in determining whether the bulk flow marks the limits of cosmic homogeneity or an early harbinger of its breakdown.
\\
\\
\noindent{\bf Acknowledgements:} 
RW was partially supported by NSF grant AST-1907365.  HAF was partially supported by NSF grant AST-1907404.  


\bibliographystyle{mnras}
\bibliography{haf}

\end{document}